# Re-establishing Kepler's first two laws for planets from the non-stationary Earth

W Y Hsiang[1], H C Chang[2], H Yao[3] and P S Lee[3]

[1]Department of Mathematics, University of California, Berkeley, CA 94720, USA.

[2]Department of Mathematics, National Taiwan University, Taipei, Taiwan 106, ROC and

[3]Department of Physics, National Taiwan Normal University, Taipei, Taiwan 116, ROC

E-mail:yao@phy.ntnu.edu.tw

**Abstract**

The Earth itself is not stationary but keeps revolving, and its motion further satisfies the law of equal area according to the heliocentric doctrine. That satisfaction can be used to construct the mathematical relationships between the planet–Sun and Earth–Sun distances. The law of equal area for planets can hence be re-established naturally from the moving Earth using the observed angular speed of a planet over the Sun. Furthermore, for the periodicity of a planet to the Sun, the distance from each planet to the Sun may be expressed as an angular periodic function. By coordinating with the observed data, this periodic distance function depicts an exact elliptical path. Here, we apply relatively simple mathematical skills to illustrate the invariant forms of planetary motions and indicate the key factors used to analyze the motions in complicated planetary systems.

## 1. Introduction

Mathematical models were widely employed to describe natural phenomena during Plato's era (427–347 BC). Spherical geometry was also applied to astronomy. Geometry is a part of cosmology, and its theory represents a realization of the structure of the entire universe. Therefore, a knowledge of geometry is crucial for understanding astronomy [1].

*Almagest*, the cosmology literature written by Ptolemy (85–165 AD), firmly established the Greek trigonometry theory that was in place for more than a thousand years. He took the concept of epicycle and deferent to depict the motion of planets, which was widely accepted by the public as absolute truth at that time. Until his pursuit of the notion of mathematical harmony and symmetry as perfection, Nicolaus Copernicus (1473–1543) strongly suspected human manipulation and complexity in the epicycle, deferent, and equant. The heliocentric doctrine was established and led to the revolution of astronomy [2, 3].

Johannes Kepler (1571–1630) was deeply enlightened by Copernicus' heliocentric theory and fully supported the doctrine throughout his life. He further proposed three laws for the planets, which described the Sun as the center of the system, and established modern astronomy theory. His theories provided a concrete basis for Isaac Newton's (1642–1727) dynamics. Kepler's laws of equal areas and ellipses were published in *New Astronomy* in 1609 [4]. The book's content is extremely obscure and generally presents complicated geometry rather than simple mathematical forms.

Most researchers have rediscovered and obtained Kepler's laws of planetary motion, either from Newton's laws of motion and universal gravitation [5-7], or from the principles of the conservation of energy and angular momentum [8-10]. To emphasize Kepler's important influence, others derived, algebraically or graphically, the inverse-square law of gravitation from Kepler's first two laws [11-13]. However, very few articles have discussed how Kepler originally derived his laws of planetary motion [14-17].

In Copernicus' and Kepler's astronomical system, the Earth is no longer stationary, which makes the determination of planet position more complicated. Fortunately, the establishment of the rules for the motion of the Earth, which are the laws of equal areas and ellipses [4, 18], present the Earth as the starting point for depicting the positions of other planets. These rules have become the basis for discovering the laws for other planets. This process, in fact, was reflected in *New Astronomy*, where Kepler divided its contents into two main parts or two inequalities. Each part treated one irregularity or non-circularity for the Earth and other planets, respectively.

To reveal the fundamental spirit of Kepler's first two laws, the present study uses simple, innovative approaches, such as trigonometric functions and the law of sines, to re-establish the laws of equal areas and ellipses for planets other than the Earth. It clarifies and simplifies the development of planet laws, which were originally difficult to interpret, enabling researchers to understand the intimate relations and analytical methods among geometry, astronomy, and physics. Moreover, this study allows researchers to practically realize the plentiful insights in major scientific developments, immerse in the joy of rediscovering scientific theories by previous great scientists, and cultivate the extensive and deep scientific prospects of these theories.

## 2. The Law of Equal Areas for Mars

The period of Mars orbiting around the Sun is approximately 687 days, which indicates that Mars will return back to the same position after 687 days; the period of the Earth orbiting around the Sun is 365 days. The periods of these two planets revolving around the Sun do not have a common value and are mutually prime numbers. This suggests the corresponding Earth position is different for every Martian year, or when Mars returns to its original position. As indicated by figure 1, if $S$ represents the Sun, $M$ is Mars or the position of Mars on the next Martian year, and $E_i$ and $E_j$ are the corresponding positions of the Earth before and after one Martian year, respectively. Therefore, a square ($SE_iME_j$) can be formed by the Sun, Mars, and two other positions of the Earth. The $SE_i$ and $SE_j$ lines represent the distance between the Sun and the different positions of the Earth, $r_i$ and $r_j$, respectively; this is known as the Earth–Sun distance. The $SM$ line represents the distance between the Sun and Mars, $d$, known as the Mars–Sun distance.

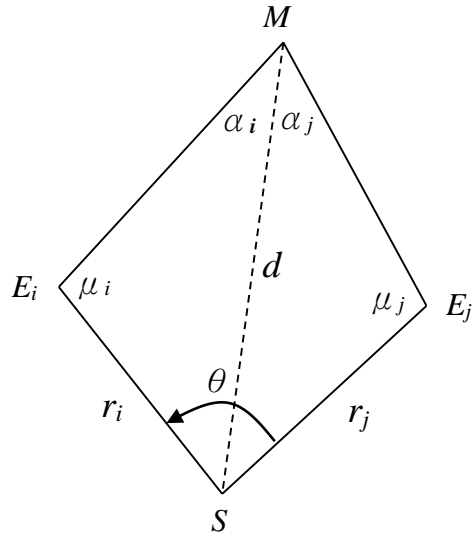

Figure 1. The illustration of the Sun (S), Mars (M), and two corresponding positions of the Earth, ($E_i$) and ($E_j$), in an adjacent Martian year.

(1) Mars–Sun Distance $d$

We randomly selected the date of $E_i$ as 5 a.m. on May 13, 1950, and the corresponding date of position $E_j$ as 4 a.m. on March 30, 1952. The time difference between $E_i$ and $E_j$ is a Martian year. The angle between S and M from the $E_i$ position is $\angle SE_iM = \mu_i$. From the observation and astronomical data from the Multiyear Interactive Computer Almanac (MICA) software [19], one found that Mars' ecliptic longitude was 172.557°, and the Sun's longitude was 51.901°, which indicated that $\mu_i$ = 172.557° − 51.901° = 120.656°. Similarly, the angle between S and M at $E_j$ was $\angle SE_jM = \mu_j$ = 9.473° + 360° − 228.333° = 141.140°.

The corresponding Sun's longitude to the Earth is the projection point of the Sun on a celestial sphere and is marked as an ecliptic longitude while observing the Sun from the Earth. The longitude is set as 0° while observing the Sun from the Earth on the vernal equinox and the Sun's longitude is set as 90° on the summer solstice as shown in figure 2. The longitudes of the Sun and any other planets can be determined from the Earth.

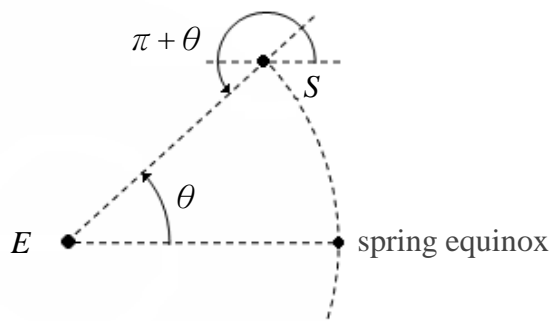

Figure 2. The Sun and planets can be directly or indirectly observed from the Earth at any moment.

The $\angle E_iSE_j = \theta$ in Fig. 1 is observable because the observed longitude of $S$ as seem from the Earth at $E_i$ and $E_j$ were 51.901° and 9.473°, respectively. Thus,

$$E_iSE_j = \theta = 51.901° - 9.473° = 42.428°. \tag{1}$$

Hence, $\theta$ can be confirmed.

The 3 observable parameters, $\mu_i$, $\mu_j$, and $\theta$, together with the law of equal areas for the Earth, will be applied hereafter to calculate the Mars–Sun distance $d$. If $\angle E_jMS = \alpha_j$, $\angle E_iMS = \alpha_i$ because the total internal angles are 360° for a quadrilateral; thus,

$$\mu_i + \mu_j + \theta + (\alpha_i + \alpha_j) = 360°, \qquad \alpha_i = 360° - \mu_i - \mu_j - \theta - \alpha_j.$$

Let $\beta = 360° - \mu_i - \mu_j - \theta$, $\beta$ is an observable value, and

$$\alpha_i = \beta - \alpha_j. \tag{2}$$

On the other hand, the quadrilateral $SE_iME_j$ can be considered as the combination of $\triangle SE_iM$ and $\triangle SE_jM$, where $SM$ is a common side. By applying the law of sines,

$$\frac{d}{\sin \mu_i} = \frac{r_i}{\sin \alpha_i}, \qquad \frac{d}{\sin \mu_j} = \frac{r_j}{\sin \alpha_j},$$

the Mars–Sun distance $d$ can be expressed as

$$d = \frac{\sin \mu_j}{\sin \alpha_j} r_j. \tag{3}$$

and the ratio between the sines of $\alpha_i$ and $\alpha_j$ may be written as follows:

$$\frac{\sin \alpha_i}{\sin \alpha_j} = \frac{r_i}{r_j} \frac{\sin \mu_i}{\sin \mu_j}, \tag{4}$$

If the law of equal areas for the Earth was established as mentioned in the first half of *New Astronomy* [4], the following relationship will hold [18]

$$r_i^2/r_j^2 = \omega_j/\omega_i.$$

This relation indicates that the ratios of the Earth–Sun distances, which were originally difficult to measure, can be calculated by the angular velocities $\omega_i$ and $\omega_j$ of the Earth at different positions, which may be measured from daily observations. Therefore, $r_i/r_j$ can be obtained. Because $\mu_i$ and $\mu_j$ are also known, the ratio in (4) can be set as an observable value $k$. Then,

$$\sin\alpha_i = k \sin\alpha_j. \tag{5}$$

Replacing (5) with (2), we have

$$\sin\beta\cos\alpha_j - \cos\beta\sin\alpha_j = k \sin\alpha_j.$$

Dividing by $\cos\alpha_j$ on both sides of the above equation, $\alpha_j$ can be found as follows:

$$\alpha_j = \tan^{-1}\left(\frac{\sin\beta}{k+\cos\beta}\right). \tag{6}$$

Hence, $\alpha_j$ turns out to be an observable value because $\beta$ and $k$ are all observable. Combining (3) and (6), the Mars–Sun distance $d$ can be directly represented by the Earth–Sun distance $r_j$. For 5 randomly selected observation dates, the Mars–Sun distances from (6) and (3) are listed in table 1. The Earth–Sun distance is set as $r_{j1} = r_0 = 100000$ on March 30, 1952 (figure 3).

Table 1. Five different randomly selected dates used to calculate the Mars–Sun distance $d$.

(The quoted and non-quoted dates have a difference of one Martian year.)

| Time | $\mu_i$ | $\mu_j$ | $\theta$ | $\beta$ | $k$ | $\alpha_j$ | $r_j$ | $d$ |
|---|---|---|---|---|---|---|---|---|
| 13 May 1950 (30 March 1952) | 120.656° | 141.140° | 42.456° | 304.253° | 1.387 | 22.973° | 100000 | 160750 |
| 21 June 1952 (9 May 1954) | 122.319° | 130.592° | 41.744° | 294.655° | 1.12 | 30.593° | 101072 | 150805 |
| 15 August 1954 (2 July 1956) | 125.251° | 116.179° | 41.593° | 283.024° | 0.906 | 40.722° | 101770 | 139993 |
| 1 November 1956 (19 September 1958) | 126.837° | 116.786° | 43.041° | 286.665° | 0.885 | 39.256° | 100518 | 141804 |
| 7 January 1959 (24 November 1960) | 122.044° | 133.781° | 44.249° | 300.073° | 1.17 | 27.383° | 98793 | 155079 |

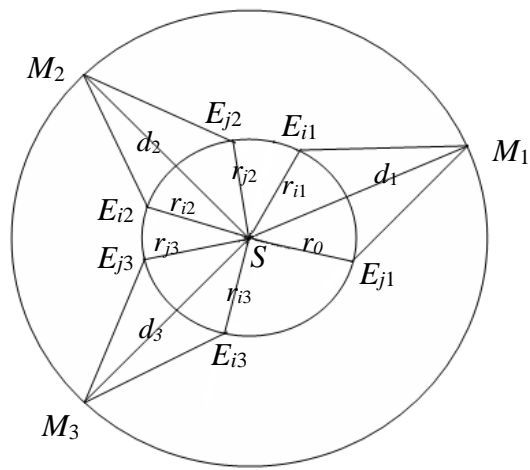

Figure 3. The Earth–Sun distance $r$ can be used to represent the Mars–Sun distance $d$.

($r_{j1} = r_0 = 100000$ is set to be a normalized value.)

## (2) Angular Speed of Mars $\omega$

The angular speed of Mars revolving around the Sun represents the angular change of Mars with respect to the Sun within a certain period of time, such as within one day. This value cannot be directly achieved by observation; the indirect relations with observable values must be determined.

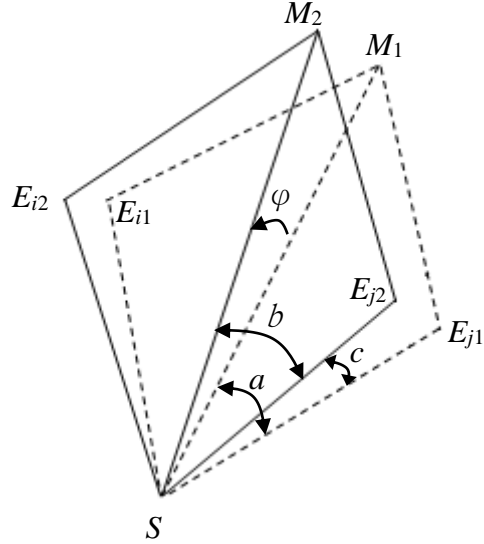

Figure 4. Two quadrilaterals $SE_{i1}M_1E_{j1}$ and $SE_{i2}M_2E_{j2}$ are formed by the Sun, Earth, and Mar within two days. The angle swept by Mars within one day from $M_1$ to $M_2$ is $\varphi = b + c - a$.

Two quadrilaterals $SE_{i1}M_1E_{j1}$ and $SE_{i2}M_2E_{j2}$ are formed by the Sun, the Earth and Mars within two days, as shown in figure 4. The value of daily angular speed of Mars $\omega$ with respect to the Sun is the value of angle $\varphi$ swept by Mars moving from $M_1$ to $M_2$. The angles $\angle SE_{i1}M_1$, $\angle SE_{j1}M_1$, and $\angle E_{i1}SE_{j1}$ in quadrilateral $SE_{i1}M_1E_{j1}$ denoted by $\mu_i$, $\mu_j$, and $\theta$ in figure 1 are also observable. The angle $\angle E_{j1}M_1S$ can further be approached from (6) like $\alpha_j$ in figure 1. As for the $\triangle SE_{j1}M_1$, using the relation of interior angles, as shown in table 1, we obtained

$$\angle M_1SE_{j1} = a = 180° - \angle SE_{j1}M_1 - \angle SM_1E_{j1} = 180° - 141.140° - 22.973° = 15.887°.$$

Similarly, $\triangle SE_{j2}M_2$ formed by the Sun, the Earth, and Mars on the second day gave

$$\angle M_2SE_{j2} = b = 180° - \angle SE_{j2}M_2 - \angle SM_2E_{j2} = 180° - 142.194° - 22.438° = 15.368°.$$

The angle $\angle E_{j1}SE_{j2}$ formed by the Sun to $E_{j1}$ and the Sun to $E_{j2}$ is similar to $\theta$ in figure 1. It could be obtained from (1) as follows:

$$\angle E_{j1}SE_{j2} = c = 10.461° - 9.473° = 0.988°.$$

Therefore, the angles swept by Mars with respect to the Sun within one day was

$$\varphi = b + c - a = 15.368° + 0.988° - 15.887° = 0.469°.$$

Furthermore, its value was the same as that of angular speed $\omega$ of Mars on that day. Table 2 lists the calculated angular velocities $\omega$ per day of Mars on five different dates from table 1.

Table 2: The calculated angular speed $\omega$ of Mars on five different dates from the observed data shown in table 1.

| Time | $\omega$ |
|---|---|
| 13 May 1950 | 0.469 |
| 21 June 1952 | 0.534 |
| 15 August 1954 | 0.620 |
| 1 November 1956 | 0.603 |
| 7 January 1959 | 0.504 |

(3) The Law of Equal Areas for Planets

The law of equal areas for a planet indicates that the line joining a planet and the Sun sweeps out an equal area in the same periods of time, as shown in figure 5, i.e.,

$$\Delta A = \frac{1}{2} d^2 \Delta \theta.$$

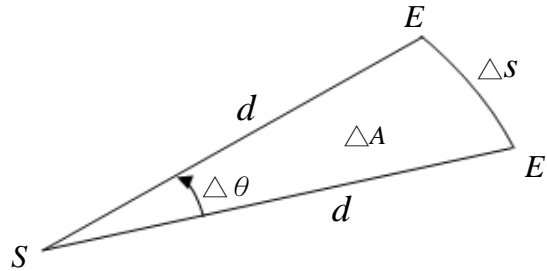

Figure 5. The law of equal areas for a planet.

Thus, the area velocity is

$$\frac{dA}{dt} = \frac{1}{2} d^2 \frac{d\theta}{dt} = \frac{1}{2} d^2 \omega,$$

where $\omega$ is the angular speed of a planet around the Sun.

Hence, to prove the law of equal areas for a planet, it is necessary to only show that the product of the square of the distance from a different planet to the Sun and the corresponding angular speed of that planet is a constant. Namely,

$$d_i^2 \omega_i = d_j^2 \omega_j. \tag{7}$$

This method is the same as that employed to inspect the equivalence of $d_j^2/d_i^2$ and $\omega_i/\omega_j$. Combining the Mars–Sun distances and angular velocities of Mars on different dates from tables 1 and 2, the corresponding values of the ratios of $d_j^2/d_i^2$ and $\omega_i/\omega_j$ can be obtained, as shown in table 3, where the referenced date is set as May 13, 1950.

Table 3. The ratios of the square of the Mars–Sun distance $d_j^2/d_i^2$ and the corresponding ratios of angular velocities $\omega_i/\omega_j$ on five different dates obtained from tables 1 and 2.

| Time | $d$ | $\omega$ | $d_j^2/d_i^2$ | $\omega_i/\omega_j$ |
|---|---|---|---|---|
| 13 May 1950 | 160750 | 0.469 | 1.000 | 1.000 |
| 21 June 1952 | 150805 | 0.534 | 0.880 | 0.878 |
| 15 August 1954 | 139993 | 0.620 | 0.758 | 0.756 |
| 1 November 1956 | 141804 | 0.603 | 0.778 | 0.778 |
| 7 January 1959 | 155079 | 0.504 | 0.931 | 0.931 |

The ratios of $d_j^2/d_i^2$ and $\omega_i/\omega_j$ for Mars are almost identical to the difference of less than 1% from the last two columns in table 3, which indicates that Mars strictly obeys Kepler's law of equal areas from the acknowledged astronomical data. This is an exciting and unsurprising result.

## 3. The Law of Ellipses for Mars

From the perspective of analytical geometry, the relationships between the Cartesian coordinates $(x, y)$ and polar coordinates $(r, \theta)$ for an ellipse with one of the foci located at $(-c, 0)$, as shown in figure 6, are

$$x = r\cos\theta - c, \quad y = r\sin\theta. \tag{8}$$

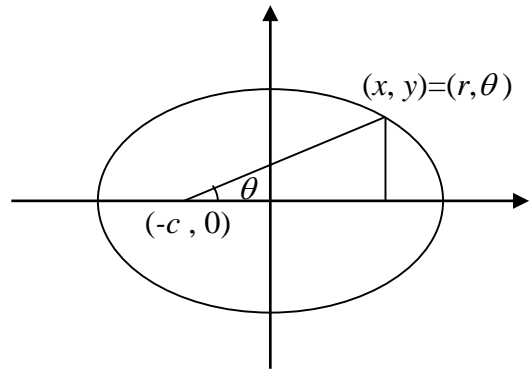

Figure 6. The relationships between the Cartesian coordinates $(x, y)$ and polar coordinates $(r, \theta)$ for an ellipse.

The equation of an ellipse in Cartesian coordinates is

$$\frac{x^2}{a^2} + \frac{y^2}{b^2} = 1 \tag{9}$$

or

$$b^2 x^2 + a^2 y^2 - a^2 b^2 = 0.$$

Substituting (9) with (8) gives

$$r(a - c\cos\theta) - b^2 = 0; \quad r(a + c\cos\theta) + b^2 = 0.$$

Taking the positive value of $r$,

$$\frac{1}{r} = \frac{a - c\cos\theta}{b^2} = \frac{a}{b^2}(1 - e\cos\theta),$$

where $e = c/a$ is the eccentricity [19]. For general situations, where $\theta \neq 0$ along the x-axis, the equation of the ellipse in polar coordinates can then be expressed as

$$\frac{1}{r} = \frac{a}{b^2}[1 - e\cos(\theta + \alpha)] = c_0 + c_1 \cos\theta + c_2 \sin\theta, \tag{10}$$

where

$$c_1^2 + c_2^2 = a^2 e^2 / b^4$$

or

$$e = b^2 \sqrt{c_1^2 + c_2^2}/a = \sqrt{c_1^2 + c_2^2}/c_0. \tag{11}$$

Therefore, (10) has the same form of ellipse as that of (9).

From the other side of the perspective, because the motion of Mars around the Sun is periodic, the distance function $d(\psi)$ from a planet to the Sun, or its reciprocal $1/d(\psi)$, can also be described as a period function of $\psi$. That is, it can be expressed as an infinite Fourier series of sines and cosines with different multiple angles as follows [20]:

$$\frac{1}{d} = a_0 + \sum_{n=1}^{\infty}[a_n \cos n\psi + b_n \sin n\psi].$$

In the ideal case, this function can be approximated by a single period of the trigonometric functions:

$$\frac{1}{d} = a_0 + a_1 \cos\psi + b_1 \sin\psi. \tag{12}$$

That is, this simplified periodic (12) is equivalent to the equation of the ellipse in polar coordinates as shown in (10).

To determine the 3 unknown $a_0$, $a_1$, and $b_1$ as shown in (12), three sets of data are required to set up simultaneous linear equations with 3 unknowns. After solving these sets of equations, the equation for the ellipse and its corresponding eccentricity can be obtained. The law of ellipses for a planet will be spontaneously revealed.

The position of Mars $M_1$ on May 13, 1950, is now selected as a reference point (figure 7; table 1). In $\triangle SE_{j1}M_1$, the angle $\angle SE_{j1}M_1 = \mu_{j1}$ is observable, and $\angle SM_1E_{j1} = \alpha_{j1}$ is calculable, which can be achieved from

(6) and is listed in table 1. Therefore, $\theta_{j1} = 180° - \mu_{j1} - \alpha_{j1} = 180.000° - 141.140° - 22.973° = 15.887°$. Similarly, $\theta_{j2} = 180.000° - 130.592° - 30.593° = 18.815°$ from the observed $\mu_{j2}$ and calculated $\alpha_{j2}$ with respect to Mars $M_2$ on June 21, 1952, from table 1. Furthermore, the angle swept by the Earth from $E_{j1}$ to $E_{j2}$ was $\angle E_{j1}SE_{j2} = 38.394°$ by the two observable angular positions of the Earth to the Sun.

Finally, the angle swept by Mars from $M_1$ to $M_2$ was obtained as $\angle M_2SM_1 = \psi = \angle E_{j1}SE_{j2} + \theta_{j2} - \theta_{j1} = 38.394° + 18.815° - 15.887° = 41.322°$, where the line connecting $M_1$ to $S$ was set to be the horizontal axis. The angles $\psi$ swept by Mars on August 15, 1954, and November 1, 1956, shown in table 1 could also be determined in a similar manner (table 4).

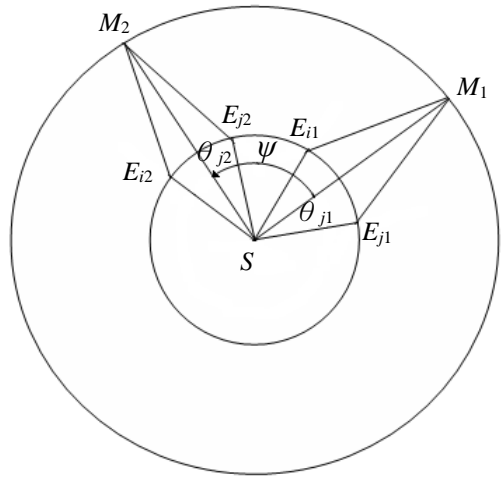

Figure 7. The angular change ($\psi$) of Mars to the Sun on the two Mars opposition

Table 4. The angle $\psi$ swept by Mars moving from the referenced position, which was selected on May 13, 1950, to the other three positions shown in table 3.

| Time | $\psi$ |
|---|---|
| 13 May 1950 | 0.000° |
| 21 June 1952 | 41.337° |
| 15 August 1954 | 98.458° |
| 1 November 1956 | 174.976° |

By replacing (12) with the Mars–Sun distances $d$ and the corresponding angles $\psi$ swept by Mars on Jun 21, 1952, August 15, 1954, and November 1, 1956, respectively, we have

$$\frac{1}{d_1} = a_0 + a_1 \cos\psi_1 + b_1 \sin\psi_1,$$

$$\frac{1}{d_2} = a_0 + a_1 \cos\psi_2 + b_1 \sin\psi_2,$$

$$\frac{1}{d_3} = a_0 + a_1 \cos\psi_3 + b_1 \sin\psi_3.$$

Solving the simultaneous linear equations in 3 unknowns, we obtain

$a_0 = 0.00000662, \quad a_1 = -0.00000040, \quad b_1 = 0.00000047.$

Thus, the periodic equation of the reciprocal of the Mars–Sun distance is

$$\frac{1}{d'} = 0.00000662 - 0.00000040\cos\psi + 0.00000047\sin\psi. \tag{13}$$

To demonstrate the generality of the above periodic equation built by 4 Mars positions, we randomly selected 5 additional dates and calculated the corresponding angles $\psi$ swept by Mars, as shown in the first two columns of table 5. The Mars–Sun distances $d'$ were estimated from periodic (13). By comparing $d'$ with $d$, which were calculated by the Mars–Sun distance (3), one could determine that $d'$ and $d$ were almost equivalent. The validity of periodic equation $1/d$ of $\psi$ in (13) can hence be asserted.

Table 5. Five randomly selected dates comparing the Mars–Sun distance $d'$ from periodic (13) with the Mars–Sun distance $d$ from (3) by the law of equal areas for the Earth.

| Time | $d$ | $\psi$ | $d'$ | $(d'-d)/d$ (%) |
|---|---|---|---|---|
| 7 January 1959 | 155090 | 235.697° | 154866 | -0.144 |
| 21 February 1961 | 164125 | 277.814° | 163935 | -0.116 |
| 25 March 1963 | 166678 | 310.942° | 166587 | -0.055 |
| 29 April 1965 | 163511 | 345.914° | 163462 | -0.030 |
| 6 June 1967 | 155039 | 24.493° | 155018 | -0.014 |

(12) or (13) has the same form as (10). Therefore, the elliptical motion of Mars revolving around the Sun as one of the foci can be verified, and the eccentricity of Mars can also be obtained by (11) as

$$e = \frac{\sqrt{c_1^2 + c_2^2}}{c_0} = \frac{\sqrt{a_1^2 + b_1^2}}{a_0} = \frac{\sqrt{(-0.00000040)^2 + (0.00000047)^2}}{0.00000662} = 0.093.$$

The result is exactly that of the well-known eccentricity of Mars, 0.093, thus again confirming the law of ellipses for Mars.

This concise method can also be applied to the other 4 planets including Jupiter, Saturn, Mercury, and Venus to determine the laws of equal areas and ellipses. Here we propose only Jupiter as an example to describe its conformity.

## 4. The Laws of Planet for Jupiter

(1) The Law of Equal Areas

Referring to figures 1 and 3 and combining (3) and (6), the Jupiter–Sun distance $d$ can also be represented by the Earth–Sun distance $r_j$, where the Earth–Sun distance is set to be $r_{j1} = r_0 = 100000$ on August 6, 1962. For the other 4 randomly selected observation dates, the calculated Jupiter–Sun distances $d$ from (3) and (6) are listed in table 6.

Table 6. Five randomly selected dates used to calculate the Jupiter–Sun distance $d$.
(Each date has one Jupiter orbital period difference.)

| Time | $\mu_i$ | $\mu_j$ | $\theta$ | $\beta$ | $k$ | $\alpha_j$ | $r_j$ | $d$ |
|---|---|---|---|---|---|---|---|---|
| 6 August 1962 | 146.236° | 152.523° | 49.469° | 348.228° | 1.191 | 5.372° | 100000 | 492868 |
| 19 October 1964 | 145.966° | 151.732° | 50.540° | 348.238° | 1.169 | 5.422° | 98199 | 492182 |
| 28 December 1966 | 144.850° | 153.833° | 50.363° | 349.046° | 1.311 | 4.737° | 96988 | 517905 |
| 26 February 1969 | 145.841° | 154.098° | 49.561° | 349.500° | 1.303 | 4.557° | 97683 | 537089 |
| 29 April 1971 | 146.757° | 153.745° | 48.787° | 349.289° | 1.251 | 4.757° | 99306 | 529690 |

Applying the relationships as shown in figure 4, one may obtain the angular speed $\omega$ of Jupiter at the different dates shown in table 6. By verifying whether the product $d^2\omega$ is a constant as shown in (7), or whether the identity $d_j^2/d_i^2 = \omega_i/\omega_j$ holds (table 7), we can establish the law of equal areas for Jupiter. From table 7, the law of equal areas for Jupiter can be certified.

Table 7. The ratios of the square of Jupiter–Sun distance $d_j^2/d_i^2$ and the corresponding ratios of angular speed $\omega_i/\omega_j$ on five randomly selected dates are nearly identical, which affirms the law of equal areas for Jupiter.

| Time | $d$ | $\omega$ | $d_j^2/d_i^2$ | $\omega_i/\omega_j$ |
|---|---|---|---|---|
| 6 August 1962 | 492868 | 0.090 | 1.000 | 1.000 |
| 19 October 1964 | 492182 | 0.090 | 0.997 | 0.997 |
| 28 December 1966 | 517905 | 0.081 | 1.104 | 1.104 |
| 26 February 1969 | 537089 | 0.076 | 1.187 | 1.188 |
| 29 April 1971 | 529690 | 0.078 | 1.155 | 1.155 |

(2) The Law of Ellipses

After the position of Jupiter on August 6, 1962 was selected as a reference point in figure 7, the angles $\psi$ swept by Jupiter on October 19, 1964, December 28, 1966, and February 26, 1969 shown in table 7 could be determined by the same method as that employed in the case of Mars. The results are listed in table 8.

Table 8. The angle $\psi$ swept by Jupiter moving from the referenced position selected on August 6, 1962, to the other three positions shown in table 7.

| Time | $\psi$ |
|---|---|
| 6 August 1962 | 0.000° |
| 19 October 1964 | 73.328° |
| 28 December 1966 | 142.104° |
| 26 February 1969 | 203.869° |

By replacing (12) with the Jupiter–Sun distances $d$ and the corresponding angles $\psi$ swept by Jupiter on October 19, 1964, December 28, 1966, and February 26, 1969, respectively, we obtain simultaneous linear equations with 3 unknowns. By solving them, the periodic equation of the reciprocal of the Jupiter–Sun distance can then be achieved as

$$\frac{1}{d'} = 0.000001950 + 0.000000075 \cos \psi + 0.000000058 \sin \psi. \qquad (14)$$

By comparing the Jupiter–Sun distances $d'$ estimated from periodic (14) with $d$ calculated by the Mars–Sun distance (3) at 5 different selected dates, we can determine that $d'$ and $d$ are nearly the same as shown in table 9. Thus, the validity of periodic equation $1/d$ of $\psi$ in (14) can be confirmed, and the elliptical motion of Jupiter revolving around the Sun as a focus may also be asserted by combining (10) and (14).

Table 9. Five randomly selected dates used to compare the Jupiter–Sun distance $d'$ from periodic (14) with the Jupiter–Sun distance $d$ from (3) by the law of equal areas for the Earth.

| Time | $d$ | $\psi$ | $d'$ | $(d' - d)/d$ (%) |
|---|---|---|---|---|
| 29 April 1971 | 529690 | 264.145° | 530717 | 0.19 |
| 5 July 1973 | 502997 | 329.491° | 503765 | 0.15 |
| 19 September 1975 | 488199 | 41.836° | 489020 | 0.17 |
| 29 November 1977 | 505574 | 113.418° | 506652 | 0.21 |
| 29 January 1980 | 531252 | 177.614° | 532653 | 0.26 |

The eccentricity of Jupiter can be calculated by (11) as

$$e = \frac{\sqrt{a_1^2 + b_1^2}}{a_0} = \frac{\sqrt{(-0.000000075)^2 + (0.000000058)^2}}{0.000001950} = 0.049.$$

The result is nearly the same as the well-known eccentricity of Jupiter, 0.048, thus again verifies the law of ellipses for Jupiter.

## 5. Conclusions

In this study, we treat the Earth as a reference point to determine the law of motions for the other planets. The fact that the Earth has regular motion, which fulfills the law of equal area, enables us to establish the mathematical relation of planet–Sun distance and Earth–Sun distance, as shown in (3). The laws of equal areas for the other planets can be easily and naturally constructed by combining this relation with the angular speed of the planet around the Sun.

The periodicity of the planet around the Sun indicates that the planet–Sun distance can be represented as the periodic function of an angle. The angular position of the observing planet and the law of equal areas are used to determine the distance of the planet to the Sun and to build the periodic function of each planet. The trajectory equation of planet distance may thus be obtained. The planet orbits are proved to be ellipses, which take the Sun as a focus; thus, the law of ellipses for planets is rediscovered.

We have applied relatively simple geometry, trigonometry, and basic algebra to describe the invariant properties of planetary motion. These procedures allow researchers to comprehend the magnitude of the mathematical approaches for analyzing the complicated and substantial planetary system, thus enabling an appreciation of the harmony and simplicity behind the natural phenomenon. The actual examples in this paper may be used by young students to establish and extend the essence and confidence applied toward scientific research.

**References**

1. Kline M 1990 *Mathematical Thought from Ancient to Modern Times*, Vol. 1 (New York: Oxford University Press)
2. Copernicus N [1543] 1995 *On the Revolutions of Heavenly Spheres* (New York: Prometheus)
3. Kuhn T 1976 *The Copernican Revolution* (Cambridge: Harvard University Press)
4. Kepler J [1609] 1992 *New Astronomy* (New York: Cambridge University Press)
5. Rainwater J and Weinstock R 1979 Inverse-square orbits: a geometric approach *Am. J. Phys.* **47** 223
6. Provost J and Bracco C 2009 A simple derivation of Kepler's laws without solving differential equations *Eur. J. Phys.* **30** 581
7. Xu D 2012 Simple derivations of Kepler's first law: use of complex variables *Eur. J. Phys.* **30** 581
8. Motz L 1975 The conservation principles and Kepler's laws of planetary motion *Am. J. Phys.* **43** 575
9. Shore F 1987 The Kepler problem recast: use of a transverse velocity transformation and the invariant velocities *Am. J. Phys.* **55**, 139
10. Noll E 2002 Teaching Kepler's laws as more than empirical statements *Phys. Educ.* **37** 245
11. Baez A 1960 Graphical derivation of the inverse-square law of gravitation from an elliptic orbit and Kepler's law of areas *Am. J. Phys.* **28** 254
12. Macklin P 1971 Inverse square law gravitation from Kepler's first and second laws *Am. J. Phys.* **39** 1088
13. Pozzi G 1977 Inverse-square gravitation from Kepler's first two laws: a Cartesian coordinate treatment *Am. J. Phys.* **45** 307
14. Wilson C 1968 Kepler's derivation of the elliptical path *Isis* **59** 5-25.
15. Aiton J 1969 Kepler's second law of planetary motion *Isis* **60** 75-90.
16. Stephenson B 1987 *Kepler's physical astronomy* (New York: Springer-Verlag Inc)
17. Voelkel J 2001 *The composition of Kepler's astronomia nova* (Princeton, NJ: Princeton University Press)
18. Hsiang WY Chang HC Yao H and Chen PJ 2011 An alternative way to achieve Kepler's laws of equal areas and ellipses for the Earth *Eur. J. of Phys*. **32** 1405-1412.
19. US Naval Observatory 2010 *Multiyear Interactive Computer Almanac* 1800-2050 (MICA)
20. Symon R 1971 *Mechanics* 3rd edn (New York: Addison-Wesley)
21. Knopp K 1996 *Theory of Functions* (New York: Dover)